\documentstyle[a4p,12pt,epsf]{article}
\def\parc{\hspace*{0.25in}}
\def\sparc{\hspace*{0.15in}}
\def\vdij{\delta{\rm\bf r}_{\rm ij}}
\def\vdrg{\delta{\rm\bf r}_{\rm rg}}
\def\vdgb{\delta{\rm\bf r}_{\rm gb}}
\def\vdbr{\delta{\rm\bf r}_{\rm br}}
\def\vn{{\rm\bf n}}
\def\vri{{\rm\bf r}_i}
\def\vrj{{\rm\bf r}_j}
\def\vrij{{\rm\bf r}_{ij}}
\def\vx{{\rm\bf x}}

\def\vrr{{\rm\bf r}_r}
\def\vrg{{\rm\bf r}_g}
\def\vrb{{\rm\bf r}_b}
\def\rrg{{\rm r}_{rg}}
\def\rgb{{\rm r}_{gb}}
\def\rbr{{\rm r}_{br}}
\def\vrrg{{\rm\bf r}_{rg}}
\def\vrgb{{\rm\bf r}_{gb}}
\def\vrbr{{\rm\bf r}_{br}}
\def\grad{\vec{\nabla}}
\def\figb{\begin{minipage}{10cm}
       \begin{center}
       \mbox{\epsfxsize=10cm\epsffile{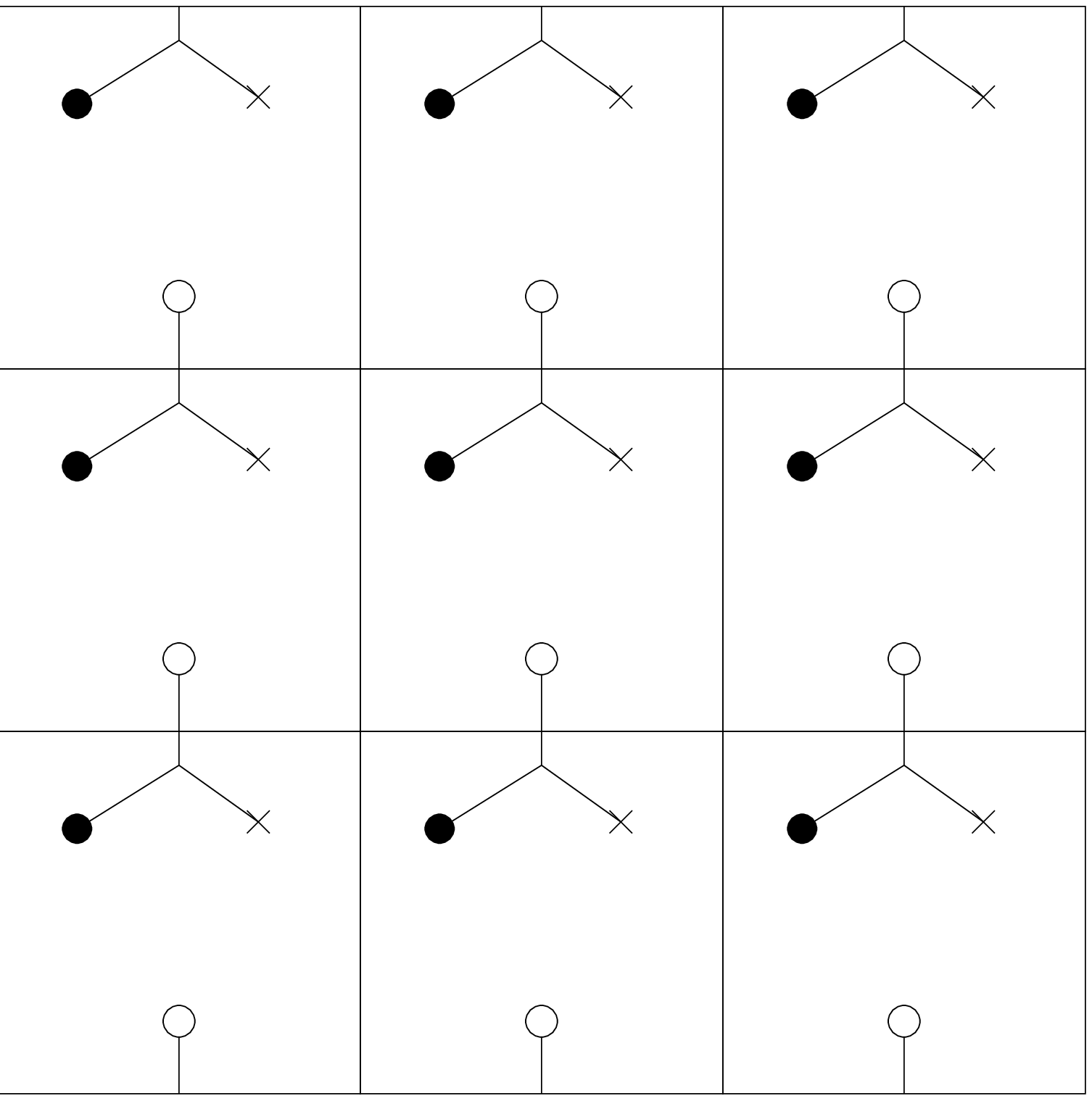}}\\\mbox{}\\
       \parbox{10cm}{\footnotesize {\bf Figure 2:} A 2-D slice showing a
       typical flux tube arrangement for three quarks, placed inside a
       central cube, subjected to periodic boundary conditions.}\\
       \end{center}
       \end{minipage}}
\def\figc{\begin{minipage}{15cm}
       \begin{center}
       \mbox{\epsfxsize=15cm\epsffile{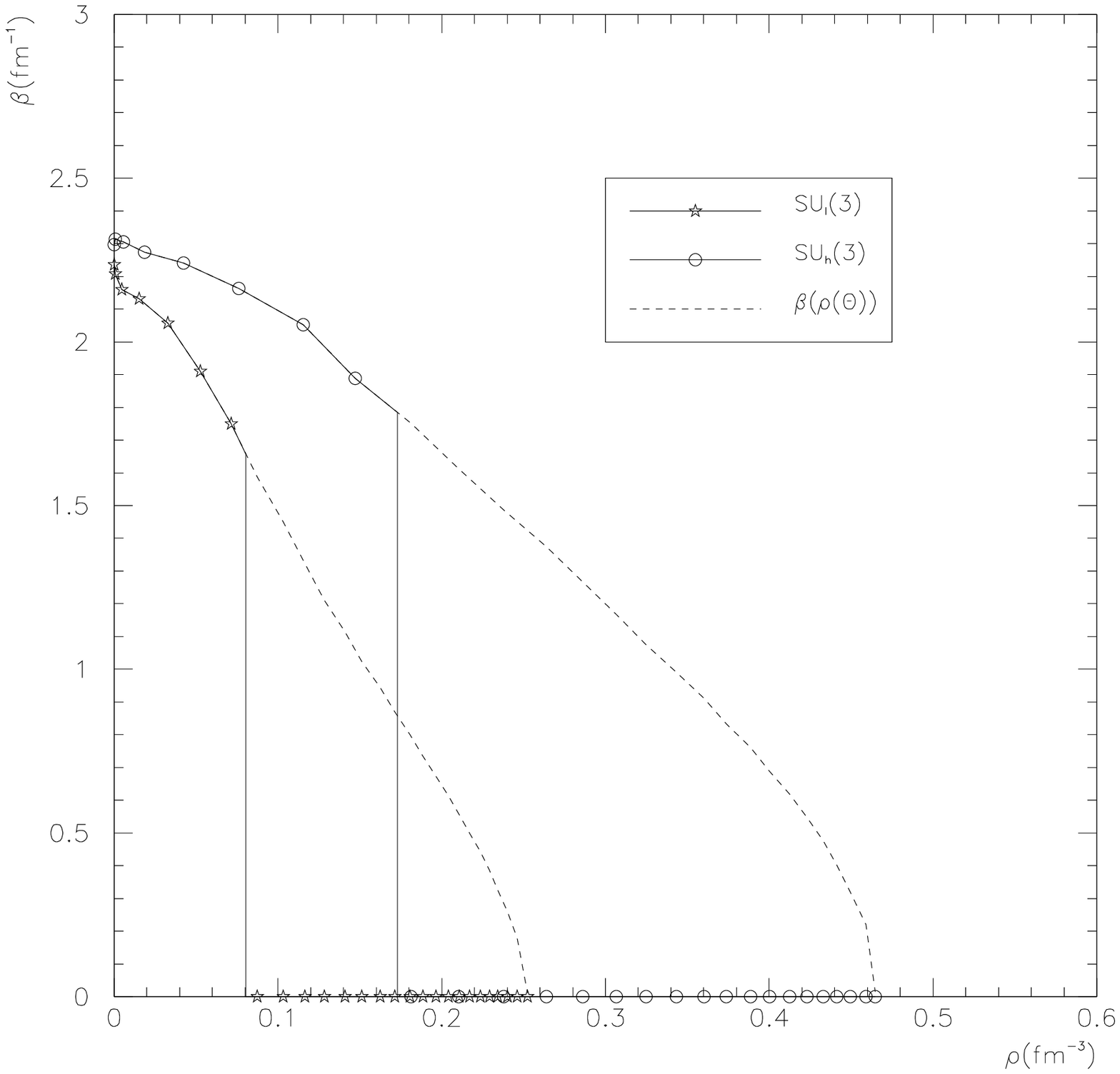}}\\
       {\footnotesize {\bf Figure 3:} Graph of $\beta(\rho)$
       for $SU_\ell(3)$ and $SU_h(3)\,$.}\\
       \end{center}
       \end{minipage}}
\def\figd{\begin{minipage}{15cm}
       \begin{center}
       \mbox{\epsfxsize=15cm\epsffile{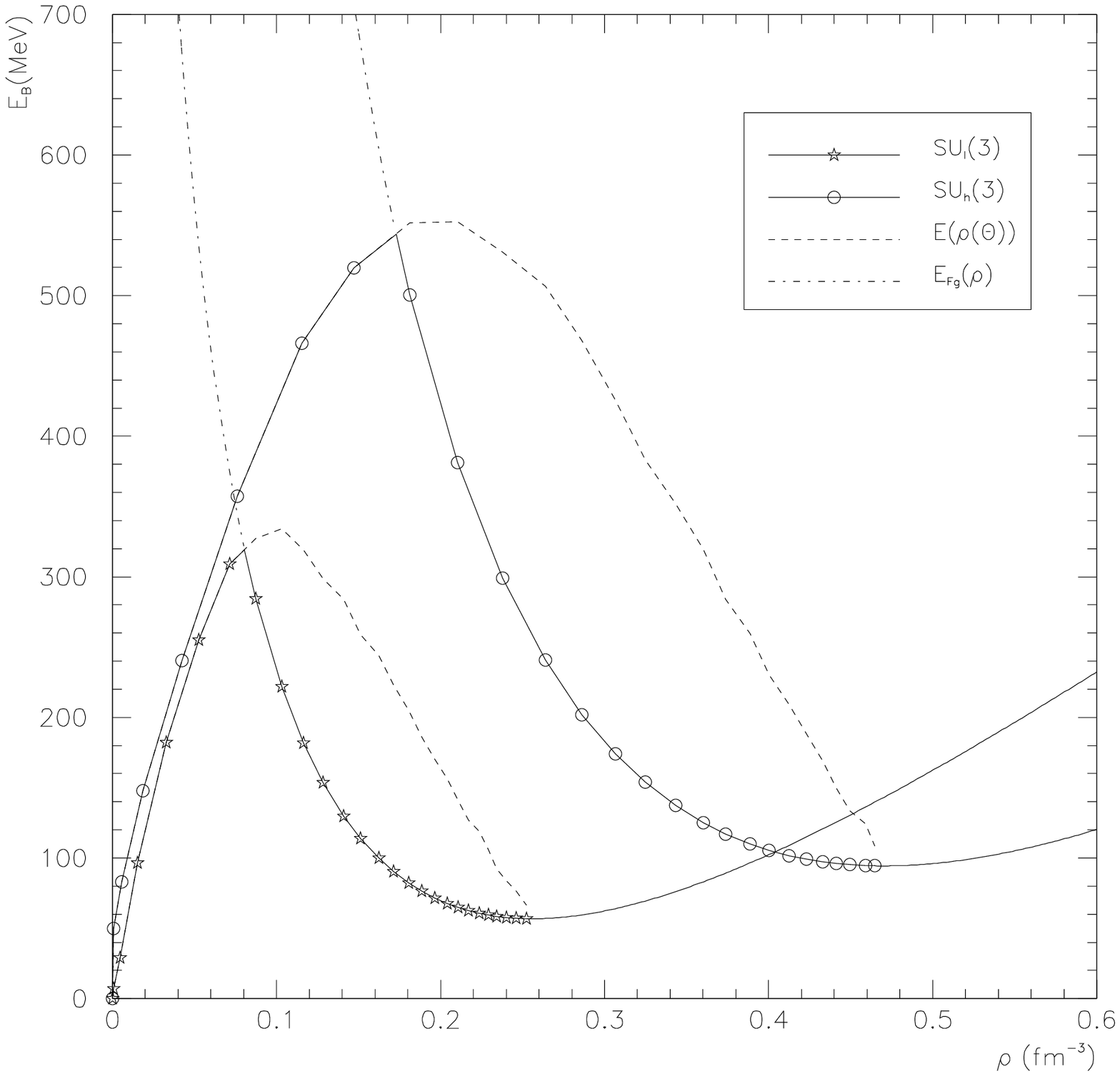}}\\
       {\footnotesize {\bf Figure 4:} Graph of $E_B(\rho)$
       for $SU_\ell(3)$ and $SU_h(3)\,$.}\\
       \end{center}
       \end{minipage}}
\def\fige{\begin{minipage}{15cm}
       \begin{center}
       \mbox{\epsfxsize=15cm\epsffile{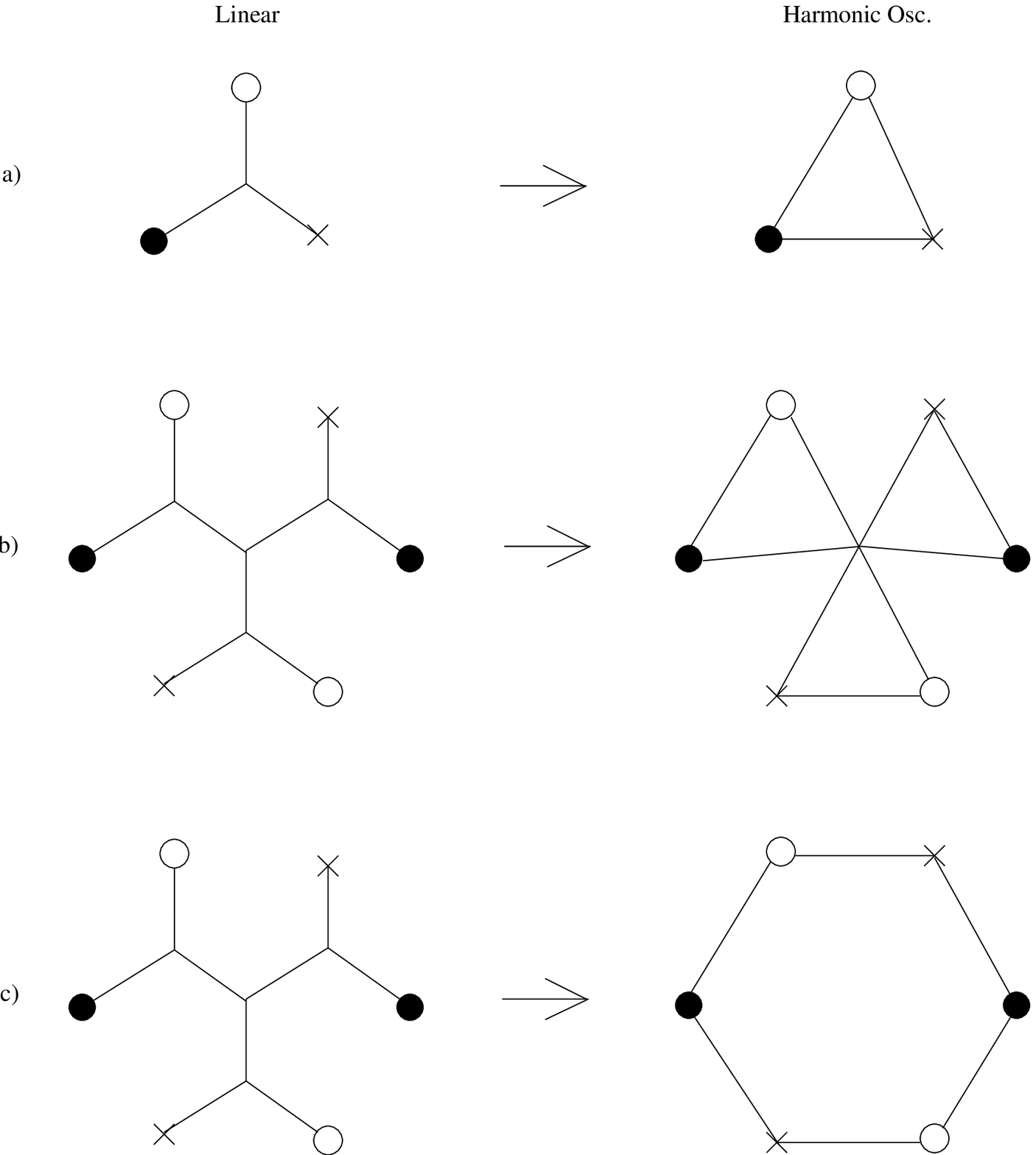}}\\\mbox{}\\
       \parbox{15cm}{\footnotesize {\bf Figure 5:} Different $SU_h(3)$
        flux tube construction schemes (RHS) motivated by their corresponding
        linear cousins (LHS). Figures a) and c) represented the HP
        construction. Figures a) and b) show a construction scheme with a
        more consistent weighting for $s$-states.}\\
       \end{center}
       \end{minipage}}
\def\figg{\begin{minipage}{10cm}
       \begin{center}
       \mbox{\epsfxsize=10cm\epsffile{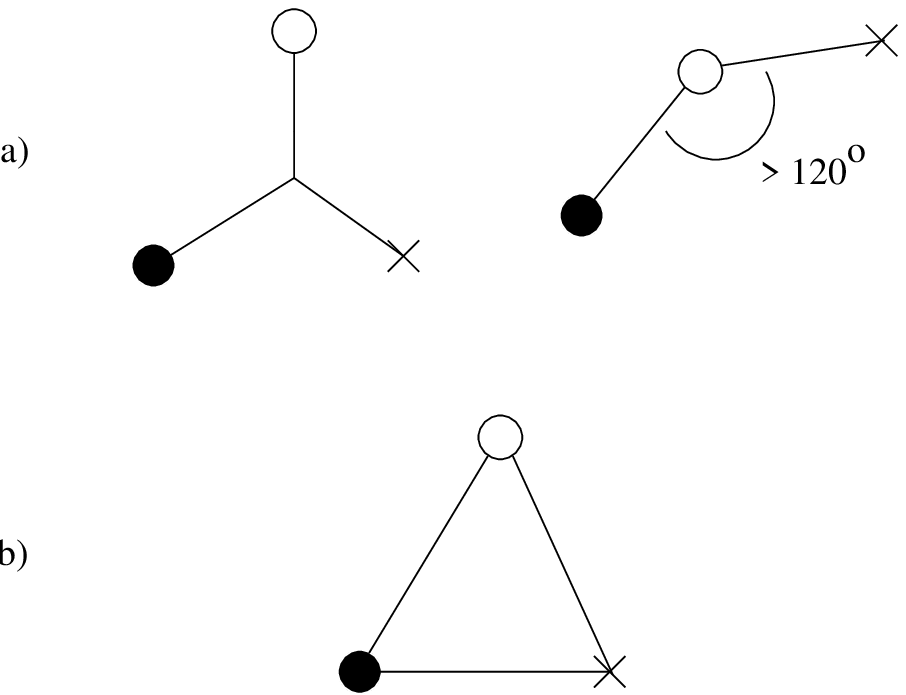}}\\\mbox{}\\
       \parbox{10cm}{\footnotesize {\bf Figure 1:} Flux tube arrangements for
       the 3q cluster potentials a) ${\rm v}_\ell$ and b) ${\rm v}_h$.}\\
       \end{center}
       \end{minipage}}
\begin{document}
\mbox{}\hfill OCIP/C-93-17\\
\mbox{}\hfill hep-ph/9312328\\
\begin{center}
{\Large\bf SU(3) String-Flip Potential Models and Nuclear Matter} \\\mbox{}\\
{\bf M. Boyce and P.J.S. Watson} \\\mbox{}\\
{\small\it Ottawa-Carleton Institute for Physics, Carleton University,\\
Ottawa, Ontario, Canada, K1S-5B6} \\\mbox{}\\
\today \\\mbox{}\\
{\bf Abstract}\\\mbox{}\\
\parbox{5.25in}{\footnotesize $\sparc$
A Monte Carlo model for nuclear matter
using a many body $SU_c(3)$ string flip potential, with fixed colour, is
investigated. The potential is approximated by considering colour singlet flux
tube formations that connect only three quarks at a time. The model is compared
with a similar string flip model, proposed by Horowitz and Piekarewicz
\cite{kn:HorowitzI}, that approximates higher order flux tube formations by
connecting quarks in colour singlet chains. The former model gives an EMC
nucleon ``swelling'' effect, whereas the latter gives an opposite effect.
Possible discrepancies between the two models are discussed.}\\
\end{center}
\begin{center}
{\bf 1. Introduction}\\
\end{center}
$\parc$There is, even after some 50 years of work on the problem, no
satisfactory explanation of the properties of nuclear matter in terms of
constituent nucleons. In principle, if QCD is the correct theory, the various
properties of nuclear matter should be calculable in a precise fashion from a
knowledge of the interactions between quarks. However, the only rigorous method
is lattice QCD, which is computationally so complex that it cannot handle more
than a single hadron. Hence we are forced to construct phenomenological models
which satisfy most of our beliefs about the interactions between quarks but are
sufficiently simple that they can be solved in a finite time.\\
$\parc$An ideal model would have the following properties: at low densities the
quarks would condense out to form isolated baryons. At a higher density, the
interaction between quarks would lead to positive binding energy between
nucleons, and a swelling of nucleons. At still higher densities, it is expected
that the hadrons will dissolve into a quark-gluon plasma. This last is in
contrast to the traditional nucleon models, which require the forces to be
carefully adjusted so that they saturate at infinite density, effectively
implying a hard core. Some simple models which appear to be likely candidates
are string-flip potential models \cite{kn:HorowitzI,kn:Watson}, and to some
extent linked cluster expansion models \cite{kn:Nzar}.\\
$\parc$The cluster models are based on one-gluon exchange potentials, and use
an N-body harmonic oscillator potential, i.e.
\begin{equation}
V_{\rm conf}=\frac{1}{2}k\sum_{i<j}(\vri-\vrj)^2\,,
\end{equation}
to mimic quark confinement. These models are mainly used for describing short
range nuclear effects, as they suffer from van der Waals forces due to the
nature of the confining potential. Despite this shortcoming, they do seem quite
useful in explaining local effects such as nucleon swelling (fat nucleons)
\cite{kn:Arifuzzaman}, quark clustering preferences, and relative strengths of
the various one-gluon exchange potentials \cite{kn:Nzar}.\\
$\parc$The string-flip potential models are, on the other hand, motivated  by
lattice QCD. They attempt to guess at how flux tubes should form amongst the
quarks at zero temperature. An adiabatic assumption is made, in which the
quarks move slowly enough for their fields to reconfigure themselves, such that
the overall potential energy is minimized: i.e.
\begin{equation}
V={\rm min}\{
           \sum_{\{q_m\ldots q_n\}}{\rm v}({\rm\bf r}_m\ldots {\rm\bf
r}_n)\,|\,
           \bigcup_{\{m\ldots n\}}^\sim\{q_m\ldots q_n\}=\{q_1\ldots q_{N_q}\}
           \}\;,
\label{eq:cpot}
\end{equation}
where the $N_q$ quarks are placed in a cube of side $L$ and subjected to
periodic boundary conditions, to simulate continuous quark matter. The sum is
over all gauge invariant sets $\{q_m\ldots q_n\}$ of quarks, such that at least
one element from each set lies inside a common box, whose disjoint union,
$\stackrel{\footnotesize\sim}{\cup}\,$, makes up the complete colour singlet
set $\{q_1\ldots q_{N_q}\}$ of $N_q$ quarks. It is easy to see that this
potential allows for complete minimal quark clustering separability at low
densities without suffering from van der Waals forces. At present these models
\cite{kn:HorowitzI,kn:Watson} are quite crude in that they do not include short
range one gluon exchange phenomena and spin effects, and are flavour
degenerate.  Despite this they do, in general, seem quite capable of getting
most of the bulk nuclear properties correct, with the exception of nuclear
binding.\\
$\parc$It is known that the $SU(2)$ string-flip potential
models do show these properties, except for the positive binding energy which
probably arises from short range forces. However, the only extension to an
$SU(3)$ model \cite{kn:HorowitzI} leads to the rather surprising result that
the nucleon appears to shrink in nuclear matter. It is therefore of some
interest to repeat the calculation of \cite{kn:HorowitzI}, in an attempt to see
whether the approximations made there alter the solution qualitatively.\\
$\parc$In this paper we construct a string-flip potential model for 3-quark
systems. Here some simplifying assumptions about flux tube minimization are
made, in order to reduce the Monte Carlo computation time.  We present results
for a linear potential model, $SU_\ell(3)$, and a harmonic oscillator potential
model, $SU_h(3)$, in which the colour has been fixed to a given quark. Our
results are compared with an $SU_h(3)$ model proposed by Horowitz and
Piekarewicz \cite{kn:HorowitzI} in which different simplifying assumptions,
about the minimal flux tube topology, were made. We also compare our earlier
$SU(2)$ results \cite{kn:Watson} with theirs \cite{kn:HorowitzI}. The paper
finishes with a discussion on future directions to pursue in attempting to get
bound state nuclear matter.
\begin{center}
{\bf 2. $SU(3)$ String-Flip Model}\\
\end{center}
$\parc$The string-flip model involves solving a Hamiltonian system of fermions
governed by the potential given in equation~(\ref{eq:cpot}). To solve this
system requires the use of variational Monte Carlo techniques
\cite{kn:Ceperley}. In order to compute  any observable in a finite amount of
time further assumptions about the form of the potential must be made.\\
$\parc$In this model the potential is restricted to summing over sets of colour
singlet clusters of three quarks,
\begin{equation}
V={\rm min}\{\sum_{\{q_rq_gq_b\}}{\rm v}(\vrr,\vrg,\vrb)|\,
\bigcup_{\{rgb\}}^\sim\{q_rq_gq_b\}=\{q_1\ldots q_{N_q}\}\}\;,\label{eq:pot}
\end{equation}
such that the colour of a given quark is fixed. The first assumption does have
some validity as it has been shown, via a linked quark cluster model, that it
is energetically more favourable for $6q$ systems to dissociate into two
nucleons as a result of hyperfine interactions
\cite{kn:Nzar,kn:MaltmanA,kn:MaltmanB}. However, this is not necessarily the
case at lower densities, as the linked cluster models are unreliable here. The
second assumption, that of fixed colour, greatly reduces  the number of degrees
of freedom, and therefore reduces the chance of finding an absolute minimum. At
low densities this should not have any effect on the potential, as the system
consists of isolated nucleons. Similarly at high densities no effect is
expected, as the system consists of uncorrelated quarks. At intermediate
densities some effects might be expected, particularly around any regions in
which a phase transition might occur.\\
$$\figg$$
$\parc$For the $SU_\ell(3)$ model, the potential, V, has the components
\cite{kn:CarlsonB}
\begin{equation}
{\rm v}_\ell(\vrr,\vrg,\vrb)=\sigma
\left\{
 \begin{array}{ll}
    \rbr+\rrg & {\rm if}\;\angle\,brg\,\ge\,120^\circ\\
    \rrg+\rgb & {\rm if}\;\angle\,rgb\,\ge\,120^\circ\\
    \rgb+\rbr & {\rm if}\;\angle\,gbr\,\ge\,120^\circ\\ \\
    \frac{1}{\sqrt{2}}\sqrt{3\xi^2+\sqrt{3}A} & {\rm otherwise}
 \end{array}
\right.\;,
\end{equation}
where $\vrij=\vri-\vrj\,$, $\xi$ or
\begin{equation}
\xi_{rgb}=\frac{1}{\sqrt{3}}\sqrt{\rrg^2+\rgb^2+\rbr^2}\label{eq:xi}\,,
\end{equation}
and
\begin{equation}
A=\frac{1}{4}\sqrt{(\rrg+\rgb+\rbr)
                   (-\rrg+\rgb+\rbr)(\rrg-\rgb+\rbr)(\rrg+\rgb-\rbr)}
\end{equation}
is the area inclosed by the triangle $\triangle rgb\,$ (see figure (1.a)).  For
$SU_h(3)$ the components are
\begin{equation}
{\rm v}_h(\vrr,\vrg,\vrb)=\frac{1}{2}k\xi_{rgb}^2\;
\end{equation}
(see figure (1.b)). This potential was obtained by replacing the linear
segments of ${\rm v}_\ell$ by springs when the quarks, which were assumed to be
of equal mass, formed a triangle with interior angles less than $120^\circ\,$
(see figure (5.a)). This analogue model is expected to have similar features to
$SU_\ell(3)$ for $s$-wave $(qqq)$ states.\\
$\parc$The sum in equation (\ref{eq:pot}) can be reordered by restricting it to
run over all sets, $\{rgb\}_L$, of quark triplets contained in a central box,
such that the potential ${\rm v}(\vrr,\vrg,\vrb)$ is minimized with respect to
all possible periodic permutations of the vectors $\{\vrr,\vrg,\vrb\}$, with
the constraint that at least one of the vectors lies inside the central box:
i.e.
\begin{equation}
V={\rm min}\{\sum_{\{rgb\}_L}v(\vrr,\vrg,\vrb)\}\,,\label{eq:mpot}
\end{equation}
where
\begin{equation}
v(\vrr,\vrg,\vrb)={\rm min}\{
                    {\rm v}(\vrr+{\bf k}_rL,\vrg+{\bf k}_gL,\vrb+{\bf k}_bL)|
k_{q_a}=-1,0,1\mbox{ \& at least one }{\bf k}_q={\bf 0}\}\,.\label{eq:metric}
\end{equation}
This means that for quark triplets a search of one box deep from the central
box is required, giving a total of $27^2$ possible permutations, in order to
minimize a given ${\rm v}(\vrr,\vrg,\vrb)$ (see figure (2)). These
permutations can be reduced to $3$ by requiring that at least two sides of the
triangle $\triangle rgb\,$, formed by a given permutation of quarks, be a
minimum: i.e.
\begin{equation}
v(\vrr,\vrg,\vrb)={\rm min}\{{\rm v}(\vdrg,\vdgb,\vrbr),
                             {\rm v}(\vrrg,\vdgb,\vdbr),
                             {\rm v}(\vdrg,\vrgb,\vdbr)\}\,,
\end{equation}
where $\vdij$ is the minimum distance vector
between the points $\vri$ and $\vrj$, in a box of side $L$ with periodic
boundary conditions, which is given by
\begin{equation}
(\vdij)_a=
      \left\{
       \begin{array}{ll}
       (\vri-\vrj)_a+L&{\rm if}\;\;(\vri-\vrj)_a\;\,<\,-L/2\\
       (\vri-\vrj)_a  &{\rm if}\;|(\vri-\vrj)_a|\,<\,\;\;L/2\\
       (\vri-\vrj)_a-L&{\rm if}\;\;(\vri-\vrj)_a\;\,>\,\;\;L/2
       \end{array}
      \right.\,,
\end{equation}
where $a=x,y,z\,$.
This is exact for $SU_h(3)$; for $SU_\ell(3)$, classical Monte
Carlo shows that about $19\%$ of the events deviate from the actual answer
by $\sim\,0.3\%$, on average.\\
$$\figb$$
$\parc$The number of different elements in the set, $\{{\footnotesize\sum}v\}$,
from which the minimum must be extracted in order to get V, defined in equation
(\ref{eq:mpot}), is $(N_n!)^2$ (where $N_n=N_q/3$ is the number of nucleons in
the central box). For $N_n=7$, say, this would be $25\,401\,600$ elements!
These elements can be reduced by fragmenting the set $\{q_1\ldots q_{N_q}\}$
into smaller pieces, or subclusters, such that each element can find $N_{br}$
complementary coloured pairs of quarks that are ``closest'' to it. These
subclusters can be further fragmented, by ``softening'' the requirement that at
least $N_{br}$ complementary pairs exist: i.e. by searching for disjoint
subclusters. These subclusters are referred to as softened subclusters. The
``closeness'' of quark $q_r$ to the complementary pair $(q_gq_b)$ is defined by
the function $\delta_{r,(gb)}=v(\vrr,\vrg,\vrb)$, given in equation
(\ref{eq:metric}). The fragmented sets are thus constructed by computing an
$N_{N_n}\times N_{N_n}^2$ matrix ($\Delta$) with elements $\delta_{r,(gb)}$,
and then converting it into block diagonal form ($\Delta^d$) increasing
in size from top to bottom, by swapping rows and columns such that each block
diagonal element contains the elements of a fragmented set and all the off
block diagonal elements are set to zero. The elements of
$\{{\footnotesize\sum}v\}$ are now constructed by extracting permutations of
elements $\delta_{r,(gb)}$, from unique columns and rows, of the block diagonal
elements of $\Delta^d\,$. Further computational speed is gained by throwing
away sums that start to exceed the current minimum. In general the fragmented
sets, constructed from $\{q_1\ldots q_{N_0}\}$, are not all disjoint from one
another, and therefore the block diagonal elements of $\Delta^d$ may overlap.
The degree of overlap turns out to increase with increasing density, causing
the Monte Carlo to slow down.\\
$\parc$This fragmentation procedure, or nearest neighbour search of depth
$N_{br}$ ($cf.$ \cite{kn:Watson}), reduces computation time quite
significantly.
The cost is that rare configurations with flux tubes that stretch across the
box, or across unsoftened subclusters, that give a global minimum might be
missed. Preliminary Monte Carlo shows that the inclusion of softened
subclusters gives no noticeable change. However, for a full $(N_n!)^2$ brute
force search, doing a Monte Carlo becomes virtually impossible. A few brute
force computations of the potential were made, for particles randomly thrown
into a box, which seem to suggest that the fragmentation procedure is
good to about $1\%\,$, with $N_{br}\approx 4$. $SU_\ell(2)$ models also give
similar results\cite{kn:Watson}.\\
$\parc$The validity of the fragmentation procedure can also be argued on
physical grounds, for it is reasonable to assume that long flux tube
configurations would tend to dissociate into $q\bar q$ pairs. Therefore the
fragmentation procedure can be consider as a zeroth order $q\bar q$
approximation.\\
$\parc$The choice of variational wave function should attempt to reflect the
overall bulk properties of the system. Here the wave
function was chosen to be of the form,
\begin{equation}
\Psi(\alpha,\beta,\rho)=
{\rm e}^{-\frac{1}{2}\;
{\displaystyle \sum_{\{rgb\}}(\beta\xi_{rgb})^\alpha}}
\prod_{c\,\varepsilon\,\{rgb\}}|\Phi_{S_c}(\rho)|\;,\label{eq:waves}
\end{equation}
where $\alpha$, $\beta$, and $\rho$ ($=N_n/L^3$ s.t. $N_n=N_q/3$) are
variational parameters, $\sum_{\{rgb\}}$ is over the set of quarks $\{rgb\}$
which gives the minimal potential V, and $|\Phi_{S_c}(\rho)|$ is
a Slater determinant with elements $\phi_{ij}=\phi_i(\vrj)$, which is
composed of the plane wave states
\begin{equation}
\phi_i(\vrj)=\sin(\frac{2\pi}{{\rm L}}\,\vn_i\cdot{\rm\bf r}_j+\delta_i)\,,
\end{equation}
where $\delta_i=0$ or $\pi/2\,$,
and $(\vn_i)_a=0,\pm 1,\pm 2,\ldots$ are the components of the Fermi
energy level packing vector $\vn_i$, for particles in a cube of side $L$,
with ordinates ranging from $-L/2$ to $L/2$, subjected to periodic boundary
conditions. This particular choice of wave function mimics the overall gross
features of quark matter, by giving highly correlated behaviour at low
densities and uncorrelated behaviour at high densities.\\
$\parc$The total energy for this many body system is,
\begin{equation}
E(\alpha,\beta,\rho)=T_{-s}+V\,,
\end{equation}
where
\begin{equation}
T_{-s}=\frac{-\hbar^2}{4m_q}(\nabla^2\ln\Psi)
\end{equation}
is the kinetic energy, obtained by eliminating the surface terms from the
integral $\int\Psi^*\nabla^2\Psi\,$. Thus the many body Hamiltonian system can
be solved by varying the parameters $\alpha$, $\beta$, and $\rho$, and
evaluating the expectation values by Monte Carlo integration at each step,
until a minimum $E$ is found.
\begin{center}
{\bf 3. Monte Carlo Calculation}\\
\end{center}
$\parc$The Monte Carlo procedure uses the Metropolis algorithm
\cite{kn:Metropolis} to generate a  distribution in $|\Psi|^2$. The procedure
then is to compute the average of an observable ${\cal O}$,
\begin{equation}
\bar{\cal O}=\frac{1}{<\Psi|\Psi>}\int{\cal O}(\vx)|\Psi(\vx)|^2d\vx
          \approx\frac{1}{N}\sum_{n=N_0}^{N+N_0}{\cal O}(\vx_n)\;,
          \label{eq:monte}
\end{equation}
where the $\sum$ is taken from $N$ sequential samples of the  distribution
$|\Psi(\alpha,\beta,\rho;\vx)|^2$ (s.t. $\alpha$, $\beta$, and $\rho$ are
fixed), after $N_0$ iterations have been made. The distribution in $|\Psi|^2$
is generated as follows: from a given configuration of particles $\vx$,  change
all their positions randomly to a new position  $\vx+d\vx$, compute  the
transition probability function
\begin{equation}
\tau={\rm min}\{(|\Psi(\vx+d\vx)|/|\Psi(\vx)|)^2,1\}\;,
\end{equation}
and compare it with a random number $r\,\varepsilon\,[0,1)$ --- if $\tau > r$
then accept the move by replacing $\vx$ with $\vx+d\vx$,  otherwise reject the
move by keeping the old $\vx$, repeat the procedure until a desired
$\delta\bar{\cal O}$ level has been reached. The intial configuration of
particles, $\vx=\vx_0$, is generated by throwing them randomly into a box of
side $L$. All subsequent moves are constrained to the box, such that if a
particle randomly moves outside, its periodic image enters from the opposite
side.  This algorithm, which satisfies detailed balance, is called the
Metropolis algorithm and converges to the distribution $|\Psi|^2$ after $N_0$
moves have been made.  The value of $N_0$ is determined by the point at which
the statistical fluctuations in $\sum_n{\cal O}_n$ have become substantially
reduced. A general rule of thumb is that convergence is more rapidly achieved
if the step size, $\delta_s x$ ({\footnotesize$=|d\vx|/\sqrt{3N_q}$}), is
chosen such that, on average, $\tau\approx 1/2\,$. A natural length scale to
use, when considering an appropriate step size, is ($cf.$ \cite{kn:Watson})
\begin{equation}
\delta_s x\sim\frac{rf}{\beta+\rho^{1/3}}\,,
\end{equation}
where the constant $f\approx1/4$. This is determined by taking several
small samples from the probability distribution $|\Psi|^2$ and by restarting
the
Monte Carlo for different $f$ values, until the desired value of $\bar\tau$ is
reached. To ensure convergence in a finite amount of cpu time, particularly
at low densities, $rgb$ clusters of quarks (of radius order $\delta_s x/2$)
are thrown into the box randomly.\\
$\parc$The Monte Carlo  evaluation of the total energy,
$\bar{E}=\overline{T_{-s}+V}$ in its current form, can produce a significant
amount of error \cite{kn:Ceperley}. This can be reduced by introducing a mean
square ``pseudoforce'',
\begin{equation}
F^2=\frac{\hbar^2}{4m_qN_n}(\grad\ln\Psi)^2\,,
\end{equation}
and re-expressing the kinetic energy as
\begin{equation}
T=2T_{-s}-F^2\,.
\end{equation}
In this form, the variance of the total energy,
\begin{equation}
\bar{E}=\overline{2T_{-s}-F^2+V}\,,
\end{equation}
goes to zero as the wave function approaches an eigenstate of the
Hamiltonian.\\
$\parc$The variational wave function $\Psi$ is made up of a product of a
correlation piece, $\chi$, and a Slater piece, $\Phi$. Therefore the kinetic
energy expression can be split up into three separate terms involving pure and
mixed, correlation and Fermi energies: i.e.
\begin{equation}
\bar{T}=\bar{T}_C+\bar{T}_F+\bar{T}_{CF}
\end{equation}
The explicit forms for these terms are: the correlation energy
\begin{equation}
\bar{T}_C=\frac{\alpha\beta^\alpha}{8m_qN_n}<
\sum_{\{rgb\}}\xi_{rgb}^{\alpha-2}[\alpha(2-(\beta\xi_{rgb})^\alpha)+8]>\;,
\end{equation}
the Fermi energy
\begin{equation}
\bar{T}_F=\frac{2\pi^2}{m_qN_nL^2}\sum_{q=1}^{N_q}\vn^2_q\,,
\end{equation}
and the mixed correlation-Fermi energy
\begin{equation}
\bar{T}_{CF}=\frac{\pi\alpha\beta^\alpha}{2m_qN_nL}<
\sum_{i=1}^{N_n}
    \sum_{\{rgb\}}
      \xi_{rgb}^{\alpha-2}
       [
         \bar{\phi}_{ir}(\vrrg-\vrbr)\phi_{ir}^\prime
        +\bar{\phi}_{ig}(\vrgb-\vrrg)\phi_{ig}^\prime
        +\bar{\phi}_{ib}(\vrbr-\vrgb)\phi_{ib}^\prime
       ]\cdot\vn_i>\;,\label{eq:mixed}
\end{equation}
where $\bar{\phi}_{ij}=(\phi^T)^{-1}_{ij}\,$, and
$\phi_{ij}^\prime=\phi_i(\vrj+\frac{L}{4\vn_i^2}\vn_i)\,$. A detailed
derivation of the above expressions can be found in the appendix.\\
$\parc$The value of $\alpha$ is fixed for free nucleons at  $\rho=0$. For
the $SU_h(3)$ model $\alpha=2$, as the wave function $\Psi$ must become that of
a free 3-body harmonic oscillator. Therefore the total energy for this system
is
simply
\begin{equation}
E_{\rm free}^{(h)}(\beta)=\frac{3\hbar^2}{2m_q}\beta^2+\frac{3k}{2\beta^2}\,.
\end{equation}
Minimizing this gives,
\begin{equation}
E_0^{(h)}=3\hbar\sqrt{\frac{k}{m_q}}\,,\label{eq:spc}
\end{equation}
where $E_0^{(h)}=E_{\rm free}^{(h)}(\beta_0^{(h)})$, and
\begin{equation}
\beta_0^{(h)}=\left(\frac{m_qk}{\hbar^2}\right)^{1/4}\,.
\end{equation}
$E_0^{(h)}$ and $\beta_0^{(h)}$ can be used to check the Monte Carlo. However,
for the $SU_\ell(3)$ model such a check is not possible, as it is impossible to
find $V$ analytically at $\rho=0\,$. We find, by fitting our results, that
\begin{equation}
E_{\rm free}^{(\ell)}(\beta)=g_{\footnotesize T}\,\frac{\hbar^2}{m_q}\beta^2
                             +g_{\footnotesize V}\,\frac{\sigma}{\beta}\,,
\end{equation}
where $g_{\footnotesize T}\approx1.07$ and $g_{\footnotesize V}\approx3.09\,$.
We can also verify the virial relation $<T_\ell>=<V_\ell>/2\,$, which should
hold at all densities. A similar check can also be done for $SU_h(3)$, with the
virial relation $<T_h>=<V_h>\,$. For $SU_\ell(3)$ the parameters $\alpha$ and
$\beta_0^{(\ell)}$ can be obtained by computing $E_{\rm free}^{(\ell)}$ for
different values of $(\alpha,\beta)$ until a minimum is found.\\
$$\figc$$
$\parc$A further reduction of the variational parameters is obtained by
introducing the scaling transformation \cite{kn:Watson},
\begin{equation}
(\beta,\rho^{1/3})\;\longrightarrow\;\zeta(\cos\theta,\sin\theta)\,,
\end{equation}
where $\zeta\,>\,0$, and $\theta$ is restricted to the interval $(0,\pi/2)$.
This allows the total energy to be expressed as a polynomial in $\zeta$, which
can subsequently be minimized to eliminate $\zeta$: i.e.
$\bar{E}(\rho,\beta)$ becomes
\begin{equation}
\bar{E}(\zeta,\theta)=
\tilde{\bar{T}\,}(\theta)\zeta^2+\tilde{\bar{V}}(\theta)/\zeta^\kappa\,,
\end{equation}
such that $\bar{T}(\zeta,\theta)=\tilde{\bar{T}\,}(\theta)\zeta^2\,$,
$\bar{V}(\zeta,\theta)=\tilde{\bar{V}}(\theta)/\zeta^\kappa$ and
\begin{equation}
\kappa=\left\{
         \begin{array}{ll}
          1  & {\rm if}\;SU_\ell(3)\\
          2  & {\rm if}\;SU_h(3)
         \end{array}
       \right.\,,
\end{equation}
which can be minimized with respect to $\zeta$ to give
\begin{equation}
\bar{E}(\theta)=(\kappa+2)
\left(
  \frac{\tilde{\bar{V}}^{\,2}(\theta)
        \tilde{\bar{T}\,}^\kappa(\theta)}{4\kappa^\kappa}
\right)^\frac{1}{\kappa+2}\,,
\end{equation}
with
\begin{equation}
\zeta(\theta)=\left(

\frac{\kappa\tilde{\bar{V}}(\theta)}{2\tilde{\bar{T}\,}(\theta)}
              \right)^\frac{1}{\kappa+2}\,.
\end{equation}
$$\figd$$
Notice that the elimination of the parameter $\zeta$ is equivalent to imposing
the virial theorem, which implies
\begin{equation}
<T>=\frac{\kappa}{2}<V>\,.\label{eq:virial}
\end{equation}
$\parc$Therefore the Monte Carlo only has to be run for different $\theta$
values extracted from the ``open'' interval $(0,\pi/2)$. The end points are
obtained by taking a limit. The $\theta=0$ limit is equivalent to taking
$\rho=0\,$, which has already been discussed. The $\theta=\pi/2$ limit is
equivalent to taking $\beta=0\,$, which corresponds to an
uncorrelated Fermi gas, with energy
\begin{equation}
E_{\mbox{\tiny Fg}}(\rho)=\left(\frac{3^5\pi^4}{2}\right)^{1/3}
\frac{3\hbar^2}{5m_q}\rho^{2/3}+V_{\mbox{\rm\tiny Fg}}(\rho)\,,
\end{equation}
where
\begin{equation}
V_{\mbox{\tiny Fg}}(\rho)=
\left\{
 \begin{array}{ll}
  {\displaystyle c_\kappa\,\frac{\sigma}{\rho^{1/3}}}&{\rm for}\;SU_\ell(3)\\
\\
  {\displaystyle c_\kappa\,\frac{k}{2\rho^{2/3}}}    &{\rm for}\;SU_h(3)
 \end{array}
\right.\,,
\end{equation}
and $c_\kappa$ is obtained by a fit to the Monte Carlo in the $\theta=\pi/2$
limit. Thus the $\beta=0$ limit is described by the curve
$E_{\mbox{\tiny Fg}}(\rho)\,$.
This curve is compared with the Monte Carlo results for $\bar{E}(\rho(\theta))$
from which a minimum energy curve $\bar{E}(\rho)$ is obtained.\\
$\parc$Figures (3) and (4) show the variational Monte Carlo results for
$\beta(\rho)$ and the binding energy, $E_B(\rho)\equiv\bar{E}(\rho)-E_0\,$,
respectively. The dashed lines on these graphs show the remnants of the minimal
$\rho(\theta)$ trajectories, for $\beta$ and $E_B\,$, after a phase transition,
at $\rho=\rho_c$, from a correlated system of quarks to an uncorrelated Fermi
gas was made. The slight roughness of these lines is because the data was not
fitted. In plotting these graphs it was assumed that: $N_n=7\,$,
$m_q=330MeV\,$, $\sigma=910MeV/fm\,$, and $k\approx3244MeV/fm^2\,$. The value
of $k$ was determined by setting $E^{(h)}_0\,$ in equation (\ref{eq:spc}) to
equal $E^{(\ell)}_0\,$, in the limit $N_n=1\,$ and $\theta=0.0001\,$. The other
parameters that were determined by the Monte Carlo are given in table (1).
\begin{center}
\begin{tabular}{c}
       {\footnotesize{\bf Table 1:} Parameters Determined by Monte Carlo, with
$N_n=7\,$.}\\
\begin{tabular}{|lr|c|c|} \hline
Parameters       &           &     $SU_\ell(3)$      &      $SU_h(3)$ \\
\hline
$\alpha$         &           &       $1.75$          &       $2.00$    \\
$c_\kappa(\rho_{\mbox{\tiny Fg}},\beta_{\mbox{\tiny Fg}})$&&$0.908$&$0.347$\\
$\rho_{\mbox{\tiny Fg}}$&$(fm^{-3})$&       $0.2524$        &$0.465$     \\
$\beta_{\mbox{\tiny Fg}}$&  $(MeV)$ &       $0.01264$       &$0.01549$    \\
$E_0(\rho_0,\beta_0)$&$(MeV)$&       $1895$          &       $1856$        \\
$\rho_0$         &$(fm^{-3})$& $1.119\times10^{-11}$ & $1.211\times10^{-11}$\\
$\beta_0$        &$(fm^{-1})$&       $2.237$         &       $2.297$
\\\hline
\end{tabular}
\end{tabular}\\
\end{center}
\begin{center}
{\bf 4. Discussion}\\
\end{center}
$\parc$The parameter $\beta$ is related to the confinement scale for triplets
of quarks. Figure (3) shows that the quarks become more deconfined as $\rho$
increases, and completely deconfined beyond the phase transition point,
$\rho_c$. Thus as $\rho$ increase from $0$ to $\rho_c$ the nucleon swells
producing an EMC-like effect. Figure (11.b), of reference \cite{kn:HorowitzI},
shows a plot of $\lambda$ ($\sim\beta^2$) $vs.$ $\rho$, obtained by
Horowitz and Piekarewicz, which in general shows that the quarks become more
confined as $\rho$ approaches $\rho_c$, and completely deconfined beyond.
Therefore their model produces a result that incorrectly explains the EMC
effect \cite{kn:Close}.\\
$$\fige$$
$\parc$ The Horowitz and Piekarewicz (HP) model approximates the higher order
flux tube topologies of equation (\ref{eq:cpot}) with long harmonic oscillator
chains that close upon themselves: i.e.
\begin{equation}
V=\min\{\sum_{\{rg\}}v_{rg}\}
  +\min\{\sum_{\{gb\}}v_{gb}\}
  +\min\{\sum_{\{br\}}v_{br}\}\,,
\end{equation}
where $v_{ij}=\frac{1}{2}kr_{ij}^2\,$, with the wave function $\Psi=e^{-\lambda
V}\Phi$ ($cf.$ equation (\ref{eq:waves}) with $\alpha=2$). They have shown for
$\rho<\rho_c$ that 3-quark clusters (chains) make up more than $90\%$ of
nuclear matter, with a large remainder of these being 6-quark clusters. A
closer look at their $\lambda$ $vs.$ $\rho$ plot shows a small dip in $\lambda$
around $\rho=0.2\,$. This indicates a slight swelling of the nucleon for very
small $\rho\,$. In fact, in this density regime 3-quark clusters completely
dominate ($>99\%$). The evidence from these graphs seems to suggest that too
much weight is being given to higher order flux tube topologies at intermediate
densities.\\
$\parc$Lattice QCD shows that quarks like to cluster together via a linear
potential. However, most phenomenological models that describe isolated
hadronic matter using a harmonic oscillator potential work just as well.  As
can been seen from figures (3) and (4) the harmonic oscillator model gives the
same overall shape as the linear one. This model was motivated by replacing
each linear segment of string in a 3-quark state by a spring, with spring
constant $k$. For quarks of equal mass this reduces to a triangle of springs
(see figure (5.a)). Similarly a 6-quark state would give an object that
simplifies to three triangles with one of the tips from each meeting at a
common vertex (see figure (5.b)). The corresponding 6-quark state for the HP
model forms a closed ring which, in general, requires less energy to form (see
figure (5.c)). Thus QCD motivated models would also seem to support the
aforementioned claim, that HP are giving too much weight to higher order flux
tube topologies.\\
$\parc$For $SU_h(2)$ our model \cite{kn:Watson} agrees with the HP model
\cite{kn:HorowitzI}. The graphs look similar to those shown in figures (3) and
(4). Of course the fact that these models agree should be of no significance,
as they both have the same potential, which only looks at combinations of
$q\bar q$ pairs. Also the $SU_h(2)$ models \cite{kn:HorowitzI,kn:Watson} when
compared with $SU_\ell(2)$ \cite{kn:Watson} gives similar contrasting figures
to the ones presented here.\\
$\parc$Figure (4), along with similar figures given in references
\cite{kn:HorowitzI,kn:Watson}, show a saturation of nuclear forces as
$\rho\rightarrow\rho_c$, followed by a phase transition to quark matter at
$\rho_c$. All of these models, however, fail to give any nuclear binding below
$\rho_c$, which would seem to suggest that the flux-tube models are incapable
of giving nuclear binding. Even the $SU_h(3)$ HP model with its long chains,
which tends to underestimate the potential, indicates that this would appear
to be the case \cite{kn:HorowitzI}. HP have a 2q model \cite{kn:HorowitzI} that
would seem to suggest that even if colour were not fixed to a given quark, no
nuclear binding would occur: albeit this model is for $p$-wave (qq) states.
Thus it would appear that string-flip models, even those that include higher
order flux tube topologies, or allow the colour to move from quark to quark,
are insufficient to obtain nuclear binding. Therefore another mechanism  for
lowering the potential must be included in these models.\\
$\parc$One way is to include one-gluon exchange interactions. As suggested by
Nzar and Hoodbhoy \cite{kn:Nzar}, the most signigicant of these are the
hyperfine interactions.
Other effects such as mass and isospin are expected to negligible.
Relativistic effects are expected, in general, to only contribute to an
overall shift down in energy.\\
\begin{center}
{\bf Conclusion}\\
\end{center}
$\parc$Various string-flip potential models have been discussed in as general a
setting as possible, and have been shown to be quite capable of describing the
bulk properties of nuclear/quark matter with the exception of nuclear binding.
At low densities they give free nucleon matter and at high densities a phase
transition to free quark matter. They show an overall saturation of nuclear
forces as nucleon densities are increased. At intermediate densities these
models, with the exception of the HP $SU_h(3)$ linked chain model
\cite{kn:HorowitzI}, give an overall EMC-like swelling of the nucleon. It is
our belief that these models with one-gluon exchange effects added on should
be capable of predicting nuclear binding.\\
$\parc$This work was supported by NSERC. The Monte Carlo calculations were
performed on an 8 node DEC 5240 UNIX CPU farm, monitored by a shepherd VAX
3100,
in the Carleton University Department of Physics.
\begin{center}
{\bf Appendix}\\
\end{center}
$\parc$Given the wave function
\begin{equation}
\Psi(\alpha,\beta,\rho)=\chi(\alpha,\beta)\Phi(\rho)\,,
\end{equation}
where
\begin{equation}
\chi(\alpha,\beta)=
{\rm e}^{-\frac{1}{2}\;{\displaystyle \sum_{\{rgb\}}(\beta\xi_{rgb})^\alpha}}
\end{equation}
and
\begin{equation}
\Phi(\rho)=\prod_{c\,\varepsilon\,\{rgb\}}|\Phi_{S_c}(\rho)|
\end{equation}
are the correlation and Fermi parts respectively, the kinetic energy can be
split up thusly: into a correlation piece
$\bar{T}_C=2\bar{T}_{C_{-s}}-\overline{F^2_C}\,$, a Fermi piece
$\bar{T}_F=2\bar{T}_{F_{-s}}-\overline{F^2_F}\,$, and a mixed correlation-Fermi
piece $\bar{T}_{CF}=-2\bar{F}_{CF}\,$, where
\begin{eqnarray}
\bar{T}_{C_{-s}}&=&\frac{-1}{4N_nm_q}<\sum_q(\nabla_q^2\ln\chi)\,>\,,\\
\bar{T}_{F_{-s}}&=&\frac{-1}{4N_nm_q}<\sum_q(\nabla_q^2\ln\Phi)\,>\,,\\
\overline{F^2_C}\;\;&=&\frac{1}{2N_nm_q}<\sum_q(\grad_q\ln\chi)^2>\,,\\
\overline{F^2_F}\;\;&=&\frac{1}{2N_nm_q}<\sum_q(\grad_q\ln\Phi)^2>\,,
\end{eqnarray}
and
\begin{equation}
\bar{F}_{CF}=\frac{1}{2N_nm_q}<\sum_q(\grad_q\ln\chi)\cdot(\grad_q\ln\Phi)>\,.
\label{eq:mcf}
\end{equation}
$\parc$The correlation pieces are straightforward to evaluate, and give the
following
\begin{eqnarray}
\grad_\ell\ln\chi
&=&-\,\frac{\alpha}{4}\,\beta^\alpha
   \sum_{\{rgb\}}\xi_{rgb}^{\alpha-2}\grad_\ell\xi_{rgb}^2\nonumber\\
&=&-\,\frac{\alpha}{6}\,\beta^\alpha\sum_{\{rgb\}}\xi_{rgb}^{\alpha-2}
   [(\vrrg-\vrbr)\delta_{r\ell}+(\vrgb-\vrrg)\delta_{g\ell}
   +(\vrbr-\vrgb)\delta_{b\ell}]\label{eq:chi}
\end{eqnarray}
with $\xi_{rgb}=\frac{1}{\sqrt{3}}\sqrt{\rrg^2+\rgb^2+\rbr^2}\,$, which yields
\begin{eqnarray}
\sum_\ell(\grad_\ell\ln\chi)^2
&=&-\,\frac{\alpha^2}{36}\,\beta^{2\alpha}\sum_{\{rgb\}}\xi_{rgb}^{2\alpha-4}
   [(\vrrg-\vrbr)^2+(\vrgb-\vrrg)^2+(\vrbr-\vrgb)^2]\nonumber\\
&=&-\,\frac{\alpha^2}{4}\,\beta^{2\alpha}\sum_{\{rgb\}}\xi_{rgb}^{2\alpha-2}\,,
\end{eqnarray}
and
\begin{eqnarray}
\sum_\ell(\nabla_\ell^2\ln\chi)
&=&-\,\frac{\alpha}{6}\,\beta^\alpha\sum_\ell\sum_{\{rgb\}}
   [\mbox{$\frac{\alpha-2}{2}$}\,\xi_{rgb}^{\alpha-4}(\grad_\ell\xi_{rgb}^2)
   +\xi_{rgb}^{\alpha-2}\grad_\ell]\nonumber\\
   &&\cdot[(\vrrg-\vrbr)\delta_{r\ell}+(\vrgb-\vrrg)\delta_{g\ell}
   +(\vrbr-\vrgb)\delta_{b\ell}]\nonumber\\
&=&-\,\frac{\alpha}{6}\,\beta^\alpha\sum_{\{rgb\}}
   \{\mbox{$\frac{\alpha-2}{3}$}\,\xi_{rgb}^{-2}
   [(\vrrg-\vrbr)^2+(\vrgb-\vrrg)^2+(\vrbr-\vrgb)^2]
   +18\}\xi_{rgb}^{\alpha-2}\nonumber\\
&=&-\,\frac{\alpha(\alpha+4)}{2}\,\beta^\alpha\sum_{\{rgb\}}
   \xi_{rgb}^{\alpha-2}\,,
\end{eqnarray}
which imply
\begin{equation}
T_{C_{-s}}=\frac{\alpha(\alpha+4)\beta^\alpha}{8m_q}
\sum_{\{rgb\}}\xi_{rgb}^{\alpha-2}\,,
\end{equation}
and
\begin{equation}
F^2_C=\frac{\alpha^2\beta^{2\alpha}}{8m_q}
\sum_{rgb}\xi_{rgb}^{2\alpha-2}\,,
\end{equation}
thus giving the desired result, i.e.
\begin{equation}
T_C=2T_{C_{-s}}-F_C^2=\frac{\alpha\beta^\alpha}{8m_q}
\sum_{\{rgb\}}\xi_{rgb}^{\alpha-2}[\alpha(2-(\beta\xi_{rgb})^\alpha)+8]\;.
\end{equation}
$\parc$The Fermi pieces are also straight forward once the following identities
are realized \cite{kn:Watson}:
\begin{eqnarray}
\grad_\ell\ln|\Phi_{S_c}|&=&
\sum_{ij}\bar{\phi}_{ij}\grad_\ell\phi_{ij}\,,\label{eq:phi}\\
\grad_\ell\bar{\phi}_{ij}&=&
-\sum_{mn}\bar{\phi}_{im}(\grad_\ell\phi_{ij}^T)\bar{\phi}_{nj}\,,\\
\nabla_\ell^2\ln|\Phi_{S_c}|&=&\sum_{ij}[\bar{\phi}_{ij}\nabla_\ell^2\phi_{ij}
-\sum_{mn}\bar{\phi}_{im}(\grad_\ell\phi_{mn}^T)
\cdot(\grad_\ell\phi_{ij})\bar{\phi}_{nj}]\,,
\end{eqnarray}
where $\bar{\phi}_{ij}\equiv(\phi^T)^{-1}_{ij}\,$, and
$\phi_{ij}=\sin(\frac{2\pi}{{\rm L}}\,\vn_i\cdot{\rm\bf r}_j+\delta_i)\,$.
These imply
\begin{equation}
\grad_\ell\phi_{ij}=\frac{2\pi}{L}\vn_i\delta_{j\ell}\phi_{ij}^\prime
\label{eq:gphi}
\end{equation}
and
\begin{equation}
\nabla_\ell^2\phi_{ij}=-\,\frac{4\pi^2}{L^2}\vn_i^2\delta_{j\ell}\phi_{ij}\,,
\end{equation}
where $\phi_{ij}^\prime\equiv\phi_i(\vrj+\frac{L}{4\vn_i^2}\vn_i)\,$.
Therefore
\begin{equation}
\sum_\ell\nabla^2_\ell\ln|\Phi_{S_c}|=-\,\frac{4\pi^2}{L^2}[\sum_{ij}
\bar\phi_{ij}\vn_i^2\phi_{ji}^T+\sum_{ijk}
\bar\phi_{ki}\phi_{ki}^\prime\vn_k\cdot\vn_j\phi_{ji}^\prime\bar\phi_{ji}]\,
\end{equation}
and
\begin{equation}
\sum_\ell(\grad_\ell\ln|\Phi_{S_c}|)^2=\frac{4\pi^2}{L^2}\sum_{ijk}
\bar\phi_{ki}\phi_{ki}^\prime\vn_k\cdot\vn_j\phi_{ji}^\prime\bar\phi_{ji}\,,
\end{equation}
imply
\begin{equation}
T_F^{(c)}=2T_{C_{-s}}^{(c)}-F_C^{2^{(c)}}
=\frac{2\pi^2}{m_qL^2}\sum_{ij}\bar\phi_{ij}\vn_i^2\phi_{ji}^T\nonumber\\
=\frac{2\pi^2}{m_qL^2}{\rm Tr}(\phi^T{\bf N}_c^2\bar\phi)\nonumber\\
=\frac{2\pi^2}{m_qL^2}{\rm Tr}({\bf N}_c^2)\,,
\end{equation}
where $({\bf N}_c^2)_{ij}\equiv\vn^2_i\delta_{ij}\,$. Thus the desired result
is
obtained by summing over the quark colour degrees of freedom, i.e.
$\sum_cT_F^{(c)}$, which implies
\begin{equation}
T_F=\frac{2\pi^2}{m_qL^2}\sum_q\vn^2_q\,.
\end{equation}
$\parc$Finally, the mixed correlation-Fermi result, given by equation
(\ref{eq:mixed}), is simply obtained, via equation (\ref{eq:mcf}), by taking
the inner product of equations (\ref{eq:chi}) and (\ref{eq:phi}), summing over
$\ell$ and $q$, and using the identity given by equation (\ref{eq:gphi}).
\vfill
\bibliographystyle{unsrt}

\end{document}